\documentclass[prl,twocolumn,floatfix,superscriptaddress]{revtex4}

\usepackage{amsmath,amssymb,amsfonts,xspace}
\usepackage{epsfig,supertabular}

\renewcommand{\Im}{\imag}

\newcommand{\be}{\begin{equation}}
\newcommand{\ee}{\end{equation}}
\newcommand{\beq}{\begin{eqnarray}}
\newcommand{\eeq}{\end{eqnarray}}
\newcommand{\bea}[2]{\be\label{#2}\begin{array}{#1}}
\newcommand{\eea}{\end{array}\ee}

\def\Im{\,{\rm Im}\, }

\def\({\left(}
\def\){\right)}
\def\[{\left[}
\def\]{\right]}

\def\11{1\!\! 1}

\def\ba{\bar a}

\def\tleta{\tilde\eta}

\newcommand{\de}{\mathrm{d}}

\newcommand{\I}{\mathrm{i}}
\newcommand{\cA}{\mathcal{A}}
\newcommand{\cB}{\mathcal{B}}
\newcommand{\cL}{\mathcal{L}}

\newcommand{\cF}{\mathcal{F}}

\newcommand{\cC}{\mathcal{C}}
\newcommand{\cS}{\mathcal{S}}

\newcommand{\cK}{\mathcal{K}}
\newcommand{\cM}{\mathcal{M}}

\newcommand{\cN}{\mathcal{N}}

\newcommand{\cX}{\mathcal{X}}

\newcommand{\cR}{\mathcal{R}}
\newcommand{\cT}{\mathcal{T}}
\newcommand{\cJ}{\mathcal{J}}

\newcommand{\cZ}{\mathcal{Z}}
\newcommand{\IR}{\mathbb{R}}

\newcommand{\IZ}{\mathbb{Z}}

\newcommand{\sk}{\mathcal{SK}}

\def\varpi{t}

\newcommand{\tzeta}{{\tilde\zeta}}

\newcommand{\kahler}{{K\"ahler}\xspace}

\newcommand{\qk}{{quaternion-K\"ahler}\xspace}

\def\bse{\begin{subequations}}
\def\ese{\end{subequations}}

\def\qli2{{\bf E}}

\def\cR#1{c(#1)}

\begin{document}

\title{
{\bf\Large
On the topology of the hypermultiplet moduli space\\Ê in
type II/CY string vacua
} }

\author{Sergei Alexandrov}
\email{salexand@univ-montp2.fr}
\affiliation{Laboratoire
de Physique Th\'eorique, Astroparticules, CNRS UMR 5207,
Universit\'e Montpellier II, 34095 Montpellier Cedex 05, France}

\author{Daniel Persson}
\email{daniel.persson@itp.phys.ethz.ch}
\affiliation{Institut f\"ur Theoretische Physik,
ETH Z\"urich,  CH-8093 Z\"urich, Switzerland}

\author{Boris Pioline}
\email{pioline@lpthe.jussieu.fr}
\affiliation{Laboratoire de Physique Th\'eorique et Hautes
Energies, CNRS UMR 7589,
Universit\'e Pierre et Marie Curie,
4 place Jussieu, 75252 Paris Cedex 05, France}

\date{November 2, 2010}

\begin{abstract}
\noindent By analyzing qualitative aspects of NS5-brane instanton
corrections, we determine the topology of the hypermultiplet moduli space
$\cM_{\text{H}}$ in Calabi-Yau compactifications of type II string theories at fixed
value of the dilaton and of the Calabi-Yau metric.  Specifically, we show that
for fivebrane instanton couplings to be well-defined, translations along
the intermediate Jacobian  must induce non-trivial shifts of the Neveu-Schwarz axion
which had thus far been overlooked. As a result,  the Neveu-Schwarz axion parametrizes
the fiber of a circle bundle, isomorphic to the one in which the fivebrane partition function
is valued. In a companion paper \cite{Alexandrov:2010ca},
we go beyond the present analysis and take steps towards a quantitative
description of fivebrane  instanton corrections, using a combination of mirror symmetry,
S-duality, topological string theory  and twistor techniques.
\end{abstract}

\maketitle


  \section{Introduction}


\noindent Determining the exact, quantum corrected  low energy effective action in
type II string theories compactified on a Calabi-Yau threefold $\cX$ is a challenging problem,
with many applications at stake. Supersymmetry requires that  the moduli space
of massless scalars is locally (and, presumably, also globally)
a product $\cM_{\text{V}}\times \cM_{\text{H}}$
of the vector multiplet (VM) moduli space $\cM_{\text{V}}$
and the hypermultiplet (HM) moduli space $\cM_{\text{H}}$ \cite{Bagger:1983tt}. While the former can
be computed exactly using  classical mirror symmetry, the latter receives
non-perturbative instanton corrections  \cite{Becker:1995kb},
which, in the current formulation of string theory,
can only be determined indirectly.

Using a combination of physical arguments (T-duality, mirror symmetry and S-duality) and
known mathematical structures (twistor techniques for \qk manifolds and wall-crossing
formulae for generalized Donaldson-Thomas invariants), D-instanton corrections
to the metric on  $\cM_{\text{H}}  $ were recently expressed in terms of the generalized
Donaldson-Thomas invariants of $\cX$ \cite{Alexandrov:2008gh,Alexandrov:2009zh},
in close parallel with the description of the HM moduli space of $\cN=2$ super Yang Mills theories
on $\IR^3\times S^1$ given in \cite{Gaiotto:2008cd}.
The metric on $\cM_{\text{H}}  $, however, should also receive instanton corrections
from NS5-branes wrapped on $\cX$. Those
are expected to restore S-duality invariance \cite{RoblesLlana:2006is}, lift
the ambiguity of the D-instanton asymptotic series \cite{Pioline:2009ia}, and resolve
the singularity
of the perturbative moduli space metric \cite{Robles-Llana:2006ez}.

Part of the difficulty of including fivebrane  instantons lies in the fact
that the type IIA fivebrane  supports a self-dual 3-form flux $H$, which is
inherently quantum mechanical  \cite{Witten:1996hc,Belov:2006jd}. In particular,
$H$ cannot be simultaneously
measured on two three-cycles $\gamma,\gamma'\in H_3(\cX,\IZ)$ with
non-zero intersection product $\langle \gamma, \gamma' \rangle$.
Equivalently, the D2-brane charge is ill-defined in the presence of a fivebrane.
As a result, the partition sum  over  fluxes/D2-branes  is not simply a function
of the metric on $\cX$ and of the background three-form field $C$, but rather a section
of a certain circle bundle $\cC_{\text{NS5}}$ over the intermediate Jacobian of $\cX$
(a torus bundle over the complex structure moduli space, with fiber
$\cT=H^3(\cX,\IR)/H^3(\cX,\IZ)$ parametrized by the $C$-field).
The restriction $\cC_{\text{NS5}}|_{\cT}\equiv \cC_\Theta$
to the torus $\cT$ further depends on a choice of  ``generalized spin structure"
$\Theta$ on $\cX$ \cite{Witten:1996hc}.

On the other hand, a fivebrane  wrapped on $\cX$ is expected to correct the
metric on $\cM_{\text{H}}$ via a term schematically
of the form
\be
\delta \de s^2\vert _{\text{NS5}} \sim e^{-4\pi r -\pi \I \sigma}\, \cZ_\Theta(C)\, ,
\label{NS5coupling}
\ee
where $r$ is related to the four-dimensional string coupling via
$r=1/g_{4}^2$, $\sigma$ is the Neveu-Schwarz (NS) axion, dual
to the Kalb-Ramond two-form in four dimensions, and $\cZ_\Theta$
is the aforementioned fivebrane partition function. The consistency of
\eqref{NS5coupling} requires that $e^{\I\pi\sigma}$ and
$\cZ_\Theta$ should be valued
in the same circle bundle $\cC_{\text{NS5}}$.

In this note, by enforcing the consistency of the NS5-instanton
correction \eqref{NS5coupling}, we
determine the topology of the hypermultiplet moduli space $\cM_{\text{H}}$
at fixed (weak) value of the string coupling and fixed complex structure
on $\cX$. This result is a crucial prerequisite for a quantitative analysis
of fivebrane instanton corrections to the metric on $\cM$, which will be presented
in the companion paper \cite{Alexandrov:2010ca}. For definiteness, we focus
on the HM moduli space
in type IIA string theory compactified on a Calabi-Yau threefold $\cX$.
By mirror symmetry and T-duality,
the same considerations apply to the  HM moduli space in type IIB string theory compactified
on the mirror threefold $\hat\cX$, or to the VM moduli space in type IIA (type IIB, respectively)
compactified on $\hat\cX\times S^1$ ($\cX\times S^1$, respectively). In
\cite{Alexandrov:2010ca}, to which the reader is referred for more details and references,
we extend the schematic coupling \eqref{NS5coupling} to the case of $k>1$ fivebranes
and promote it to an actual deformation of the
metric on $\cM_{\text{H}}  $ by combining insights from mirror symmetry, S-duality,
topological string theory and  twistor techniques.

\section{ Perturbative hypermultiplet metric}

\noindent Recall that in type IIA string theory, the HM moduli space describes the
vacuum expectation values of  the dilaton $r$ and NS-axion $\sigma$,
introduced in \eqref{NS5coupling}, the complex structure of $\cX$
and the periods of the 3-form field $C$ on $H_3(\cX,\IZ)$. To write the
metric explicitly, let us choose a symplectic basis $(\mathcal{A}^\Lambda,
\mathcal{B}_\Lambda)$, $\Lambda=0,\dots, h_{2,1}(\cX)$,
of $\Gamma\equiv H_3(\cX,\mathbb{Z})$
and complex coordinates $z^a$, $a=1,\dots, h_{2,1}(\cX)$, on the
complex structure moduli space $\cM_c(\cX)$. In the one-loop
approximation,  the metric element on $\cM_{\text{H}}  $ can be written as follows
\cite{Gunther:1998sc,Robles-Llana:2006ez,Alexandrov:2007ec}:
\beq
\label{hypmetone}
\de s_{\cM}^2&=&\frac{r+2c}{r^2(r+c)}\,\de r^2
+\frac{4(r+c)}{r}\, \de s^2_{\cS\cK}
+\frac{\de s^2_\cT }{r}
\\
&+& \frac{2\, c}{r^2}\, e^{\cK}\, | X^\Lambda \de \tzeta_\Lambda - F_{\Lambda} \de \zeta^\Lambda|^2
+ \frac{r+c}{16 r^2(r+2c)} D\sigma^2  .
\nonumber
\eeq
Here, $\Omega\equiv (X^\Lambda,F_\Lambda)$ are the complex periods of the holomorphic three-form $\Omega_{3,0}$
along the symplectic basis $(\mathcal{A}^\Lambda, \mathcal{B}_\Lambda)$, $\de s^2_{\cS\cK}$
is the special \kahler metric on the complex structure moduli
space $\cM_c(\cX)$,  with K\"ahler potential
$\cK=-\log[ \I ( \bar X^\Lambda F_\Lambda - X ^\Lambda \bar F_\Lambda)]$,
$C\equiv( \zeta^\Lambda, \tzeta_\Lambda)\in \cT$ are the real periods of the $C$-field
on the symplectic basis of $H_3(\cX, \mathbb{Z})$,
 \be
\label{dsT}
\de s^2_\cT = -\frac12
(\de \tzeta_\Lambda-\cN_{\Lambda\Lambda'} \de\zeta^{\Lambda'})
 \Im \cN^{\Lambda\Sigma}
 (\de \tzeta_\Sigma-\bar\cN_{\Sigma\Sigma'} \de\zeta^{\Sigma'})
\ee
is the \kahler metric  on the intermediate Jacobian torus
$\cT=H^{3}(\mathcal{X},\mathbb{R}) / \Gamma$\
(where we identify $\Gamma $ with $ H^3(\cX,\mathbb{Z})$, neglecting torsion),
 $\mathcal{N}_{\Lambda\Sigma}$ is the period
matrix in the Weil complex structure,  related to the Griffiths
period matrix $\tau_{\Lambda\Sigma}=\partial_\Lambda \partial_\Sigma F(X)$
($F(X)$ being the prepotential) via the usual special geometry relation
\be
\label{defcN}
\cN_{\Lambda\Lambda'} =\bar \tau_{\Lambda\Lambda'} +2\I \,\frac{ [\Im\tau \cdot X]_\Lambda
[\Im \tau \cdot X]_{\Lambda'}}
{X^\Sigma \, \Im\tau_{\Sigma\Sigma'}X^{\Sigma'}}\, ,
\ee
and, finally, $D\sigma$ is the one form
\be
\label{Dsigone}
D\sigma = \de \sigma + \tzeta_\Lambda \de \zeta^\Lambda -  \zeta^\Lambda \de \tzeta_\Lambda
+ 8 c \cA_K\, ,
\ee
where $\cA_K=\frac{\I}{2}( \cK_a \de z^a -  \cK_{\ba} \de \bar z^{\ba})$ is the \kahler
connection on $\cM_c(\cX)$.
The value of the parameter $c$
was determined by a one-loop computation in \cite{Antoniadis:1997eg,Antoniadis:2003sw} to be
\be
c=-\chi(\cX)/(192 \pi) \, ,
\label{cchi}
\ee
where $\chi(\cX)$ is the Euler number of $\cX$. This value was later seen to be
consistent with S-duality on the type IIB side \cite{RoblesLlana:2006is}.
Alternatively,  \eqref{cchi} may be derived
by dimensionally reducing the topological coupling $B\wedge I_8$ in the low energy
effective action of type IIA string theory in 10 dimensions and dualizing $B$
into~$\sigma$ \cite{Alexandrov:2010ca}. The actual value of $c$ will only become important
at the end of this note.

For any value of the parameter $c$, the metric \eqref{hypmetone} is, as required
by supersymmetry \cite{Bagger:1983tt}, \qk (though not complete when $c<0$,
due to a curvature singularity at $r=-2c$), and asymptotes to the $c$-map metric \cite{Cecotti:1988qn,
Ferrara:1989ik} in the weak coupling limit $r\to \infty$. It moreover admits a
continuous group of isometries
\be
\label{heis0}
\begin{split}
T_{H,\kappa} :
\bigl(\zeta^\Lambda, \tzeta_\Lambda,\sigma\bigr)\mapsto &
\bigl(\zeta^\Lambda + \eta^\Lambda ,\tzeta_\Lambda+ \tleta_\Lambda \bigr. ,
\\ &
\bigl. \sigma + 2 \kappa - \tleta_\Lambda \zeta^\Lambda + \eta^\Lambda \tzeta_\Lambda
\bigr)
\, ,
\end{split}
\ee
with $H=(\eta^\Lambda,\tleta_\Lambda)\in \IR^{b_3(\cX)}$, $\kappa \in \IR$, satisfying
the Heisenberg group law
\be
T_{H_2,\kappa_2} T_{H_1,\kappa_1} =
T_{H_1+H_2,\kappa_1+\kappa_2+\frac{1}{2}\langle H_1, H_2\rangle}\, ,
\label{grouplaw}
\ee
where $\langle H_1, H_2 \rangle\equiv \tilde\eta_{\Lambda,1} \eta_2^\Lambda-\tilde\eta_{\Lambda,2} \eta_1^\Lambda$
is the natural symplectic pairing on $H^3(\cX,\IR)$
(for $c=0$, there is an additional continuous isometry rescaling $(r,C,\sigma)$ with weights
$(1,1,2)$ but it is broken when $c\neq 0$). In addition, the metric \eqref{hypmetone}
admits discrete isometries corresponding to monodromies in $\cM_c(\cX)$, accompanied
by an integer symplectic action on the vector $C$ and a shift of $\sigma$,
\be
\label{sigmon}
C\mapsto \rho(M)\cdot C\, ,\qquad
\sigma \mapsto \sigma + \frac{\chi(\cX)}{24\pi} \Im\, f_M \ ,
\ee
where  $f_M$ is a local holomorphic function on $\cM_c(\cX)$ determined by the rescaling
$\Omega_{3,0}\mapsto e^{f_M} \Omega_{3,0}$ undergone by the holomorphic 3-form
under  monodromy.

The metric \eqref{hypmetone} is presumed to be valid to all orders in $1/r$, by
standard non-renormali\-zation arguments (see e.g. \cite{Alexandrov:2008nk}). It does
however receive D- and
NS5-instanton corrections at order $e^{-\sqrt{r}}$ and $e^{-r}$, respectively,
which break the continuous isometries \eqref{heis0} to a discrete subgroup
corresponding to large gauge transformations. Our main
goal in this note is to identify this group.

\section{D-instanton corrections}

\noindent  In the type IIA set-up, D-instantons  originate from Euclidean D2-branes
wrapped on a special Lagrangian (sLag) submanifolds of $\cX$ (more
generally, from stable objects in the Fukaya category of $\cX$). Qualitatively,
they induce corrections to the metric \eqref{hypmetone}
of the following form (see \cite{Alexandrov:2008gh} for the precise result)
\be
\label{d2quali}
\delta \de s^2\vert_{\text{D2}} \sim \Omega(\gamma, z^a)\, \sigma_{\text{D}}(\gamma)\,
e^{ -8\pi \sqrt{r} |Z_\gamma|
- 2\pi\I (q_\Lambda \zeta^\Lambda-p^\Lambda\tzeta_\Lambda)} ,
\ee
where $(p^\Lambda,q_\Lambda)$ are integers which label the homology class
$\gamma=q_\Lambda \cA^\Lambda- p^\Lambda \cB_\Lambda\in\Gamma$,
$Z_\gamma\equiv e^{\cK/2} (q_\Lambda X^\Lambda-p^\Lambda
F_\Lambda)$ is the volume (or, in mathematical parlance, the stability data)
of any sLag in the homology class $\gamma$, $\Omega(\gamma,z^{a})$
is  Joyce's invariant \cite{MR1941627} (a particular instance of generalized Donaldson-Thomas
invariants), and $\sigma_{\text{D}}(\gamma)$ is a quadratic refinement of the symplectic
pairing on $H_3(\cX,\IZ)$, i.e. a phase assignment $
\sigma_{\text{D}}:H_3(\cX,\IZ)\rightarrow U(1)$ such that
\be
\label{qrifprop}
\sigma_{\text{D}}(\gamma+\gamma') = (-1)^{\langle \gamma, \gamma' \rangle}\,
\sigma_{\text{D}}(\gamma)\, \sigma_{\text{D}}(\gamma') \, .
\ee
As explained in \cite{Gaiotto:2008cd}, this phase factor is crucial in ensuring
consistency with the Kontsevich-Soibelman wall-crossing formula \cite{ks},
and hence  smoothness of the metric across lines of marginal stability
in $\cM_c(\cX)$ where $\Omega(\gamma, z)$ jumps. Having chosen
a Lagrangian decomposition $\Gamma=\Gamma_e \oplus \Gamma_m$
into $\cA$ and $\cB$ cycles,
the solutions of \eqref{qrifprop} can be parametrized by characteristics
$\Theta_{\text{D}}=(\theta_D^\Lambda,\phi_{D,\Lambda})\in \cT$
such that \cite{Belov:2006jd}
\be
\sigma_{\text{D}}(\gamma) = e^{-\I\pi  q_\Lambda p^\Lambda + 2\pi \I ( q_\Lambda \theta_D^\Lambda
- p^\Lambda \phi_{D,\Lambda})}\, .
\label{quadraticrefinementD}
\ee
The characteristics $\Theta_{\text{D}}$ may be set to zero by redefining
$\hat C=C-\Theta_{\text{D}}$, at the
cost of spoiling the transformation property \eqref{sigmon} of $C$ under
monodromies, as observed in \cite{Gaiotto:2008cd}.
The essential quadratic term in \eqref{quadraticrefinementD},  however,
cannot be disposed of.

Leaving the prefactor in \eqref{d2quali} aside for now, we see that the D-instanton corrections
break the continuous isometries $T_{H,\kappa}$ to a subgroup
where $H\in H^3(\cX,\IZ)$ and $\kappa$ can still take any value in $\mathbb{R}$. In particular,
keeping the dilaton fixed and quotienting out by translations along the NS-axion, the
HM moduli space reduces to the intermediate Jacobian of $\cX$ \cite{Morrison:1995yi,
Aspinwall:1998bw,StienstraVandoren}.

\section{The fivebrane partition function}

\noindent  Let us now (re)turn to the NS5-instanton coupling \eqref{NS5coupling}.
As first discussed in \cite{Witten:1996hc}, the partition function $\cZ_\Theta$ of a self-dual
three-form on the fivebrane  worldvolume $\cX$ is a section of a circle bundle
$\cC_{\text{NS5}}$ over the intermediate
Jacobian $\cJ_c(\cX)$, whose  restriction to the torus $\cT$ has
 first Chern class equal to the \kahler class,
\be
c_1(\cC_{\text{NS5}})\vert_{\cT} =  \omega_\cT \equiv \de\tzeta_\Lambda  \wedge \de\zeta^\Lambda\, .
\ee
For a  fixed metric on $\cX$, such a circle bundle over $\cT$ is determined by the holonomies
$\sigma_{\rm NS5}(H)\in U(1)$ around one-cycles in $\cT$ (equivalenly, three-cycles
$H=(\eta^\Lambda,\tilde\eta_\Lambda)\in H^3(\cX,\IZ)$), satisfying the relations
\be
\label{qrifprophol}
\sigma_{\rm NS5}(H+H') = (-1)^{\langle H,H' \rangle}\,
\sigma_{\rm NS5}(H)\, \sigma_{\rm NS5}(H') \, ,
\ee
identical in form to the defining relation \eqref{qrifprop} of a quadratic refinement.
As before, we can solve \eqref{qrifprophol} using characteristics $\Theta=(\theta,\phi)$,
so that
\be
\sigma_{\text{NS5}}(H) = e^{-\I\pi\, \tilde\eta_\Lambda \eta^\Lambda
+ 2\pi\I ( \tilde\eta_\Lambda \theta^\Lambda
- \eta^\Lambda \phi_{\Lambda})}\, .
\label{quadraticrefinement}
\ee
We shall later argue that $\sigma_{\text{D}}$ and $\sigma_{\rm NS5}$ should be chosen to be identical but
for now we keep them distinct.

Having chosen the holonomies $\sigma_{\text{NS5}}(H)$, the bundle $\cC_{\Theta}\equiv \cC_{\text{NS5}}\vert_\cT$
is now defined by the twisted periodicity property of its sections
$\cZ_\Theta (C)$ under large gauge transformations \cite{Belov:2006jd},
\be
\label{thperiod}
\cZ_\Theta (C+H)
= \sigma_{\rm NS5}(H) \, e^{\I\pi(\eta^{\Lambda}\tilde\zeta_\Lambda-\tleta_\Lambda\zeta^{\Lambda})}
\cZ_\Theta (C)\, ,
\ee
for all $H=(\eta^{\Lambda},\tilde\eta_\Lambda)\in H^3(\cX,\IZ)$.
It is easy to see that such sections can always be written as a generalized theta
series
\be
\label{thpl2f}
\cZ_\Theta(C) =
\!\! \!\!\!\sum_{n^\Lambda\in \Gamma_m+\theta}\!\!\! \!
\Psi\left( \zeta^\Lambda - n^{\Lambda}\right)
e^{2\pi\I   (\tilde\zeta_{\Lambda}-\phi_\Lambda) n^{\Lambda}
+\I\pi (\theta^{\Lambda}\phi_{\Lambda}-\zeta^{\Lambda}\tilde\zeta_{\Lambda})} ,
\ee
where the kernel $\Psi(\zeta^\Lambda)$ is an arbitrary function on $\Gamma_m\otimes \IR$,
which may in general depend on the metric of $\cX$. In particular, starting
from the partition function of a non-chiral Gaussian three-form on $\cX$ and
performing holomorphic factorization  \cite{Witten:1996hc,Henningson:1999dm,Dijkgraaf:2002ac,Belov:2006jd}
leads to a particular solution of \eqref{thperiod} given by a Gaussian kernel
\be
\label{PsiGauss}
\Psi(\zeta^\Lambda) = \cF
\exp\left(\I\pi\, \zeta^\Lambda \bar\cN_{\Lambda\Sigma} \zeta^\Sigma\right) ,
\ee
where $\cF$ is a normalization factor which depends on the complex structure of $\cX$. This
solution is proportional to the standard Siegel theta series of rank $b_3(\cX)/2$,
and has  the additional feature of being holomorphic
with respect to the Weil complex structure on $\cT$.
The exact fivebrane  partition function, which we compute in \cite{Alexandrov:2010ca}
using S-duality and twistorial techniques, is in general
non-Gaussian and non-holomorphic, but it does reduce to this solution in the
weak coupling limit. In particular, substituting \eqref{PsiGauss} in
\eqref{NS5coupling}, we recover the  classical fivebrane  instanton action
expected from the supergravity analysis of  \cite{deVroome:2006xu},
\beq
\label{SdV}
S_{\rm NS5}&=&\,\pi \left[ 4 r
- \I (n^{\Lambda}-\zeta^\Lambda)\bar \cN_{\Lambda\Sigma} (n^{\Sigma}-\zeta^\Sigma)\right]
\\
&&+ \I\pi \bigl[\sigma+\zeta^\Lambda \tzeta_\Lambda
-2 n^{\Lambda}(\tilde\zeta_{\Lambda}-\phi_\Lambda)  - \theta^\Lambda \phi_\Lambda\bigr]\, .
\nonumber
\eeq

At this stage we wish to stress two important points.
First, while the bundles $\cC_\Theta$ and $\cC_{\Theta'}$ are not isomorphic
when $\Theta-\Theta'\notin \Gamma$, the
corresponding theta series for the same kernel $\Psi(\zeta)$
are nevertheless related by a simple shift of $C$,
\be
\label{thschar}
\cZ_{\Theta}(C) =
e^{\I\pi \left( \langle \Theta,\Theta'\rangle
 +  \langle C,\Theta- \Theta'\rangle\right)}\,
\cZ_{\Theta'}(C+\Theta'-\Theta).
\ee
This observation will be relevant in Eq. \eqref{redef} below. Second, the theta series
\eqref{thpl2f} assumes a choice of Lagrangian decomposition $\Gamma=
\Gamma_e\oplus \Gamma_m$. Under
a monodromy in $\cM_c(\cX)$, this choice will generally not be preserved. The partition
function $\cZ_\Theta(C)$ will nevertheless be invariant (after the necessary transformation
of $C$, $\Theta$ and the period matrix $\cN$) provided $\Psi(\zeta)$ stays invariant under
the action of $\rho(M)$ via the metaplectic (Schr\"odinger-Weil) representation.
This indeed holds for the
Gaussian solution \eqref{PsiGauss}, and provides a strong constraint on its non-Gaussian
generalization. It also suggests that there should be a direct relation between the
exact $\Psi(\zeta)$ and the topological string wave-function, which we spell out
in \cite{Alexandrov:2010ca}.

\section{Topology of the NS-axion circle bundle $\cC$ over $\cT$}

\noindent  Having recalled some basic properties of the fivebrane  partition function $\cZ_\Theta$,
we can now analyze the implications of the instanton correction \eqref{NS5coupling}
for the topological nature of the NS-axion $\sigma$. The first, obvious observation is that
\eqref{NS5coupling} breaks continuous shift symmetries $\sigma\mapsto\sigma+2\kappa$
to those with integer $\kappa$. Thus, $e^{\I\pi\sigma}$, $0\leq \sigma<2$ parametrizes
the fiber of a certain circle bundle $\cC$ over the intermediate Jacobian $\cJ_c(\cX)$,
to be identified.

The second  observation is that
$e^{\I\pi\sigma}$ must transform in the same way
as $\cZ_\Theta$, Eq. \eqref{thperiod},  under large gauge transformations. This requires that under
$C\mapsto C+H$, $\sigma$ should simultaneously shift according to
\be
\sigma\mapsto \sigma
- \tleta_\Lambda \zeta^\Lambda+\eta^\Lambda \tzeta_\Lambda
+ 2\cR{H}\, ,
\ee
where $\cR{H}$ is defined modulo 1 by
$\sigma_{\text{NS5}}(H)=(-1)^{2\cR{H}}$,
 i.e.
\be
\label{propkappaH}
\cR{H}=- \frac12\, \eta^\Lambda \tleta_\Lambda + \tleta_\Lambda\theta^\Lambda
- \eta^\Lambda\phi_\Lambda \ \ \ {\rm mod}\   1\, .
\ee
Due to the periodicity of $\sigma$ modulo 2, the ambiguity of $\cR{H}$ modulo
the addition of integers is irrelevant.  Thus,
we conclude that  large gauge transformations are  generated by
$T'_{H,\kappa}\equiv T_{H,\kappa+\cR{H}}$ with
$H\in H^3(\cX,\IZ)$ and $\kappa\in \IZ$, acting as
\be
\label{heisext}
\begin{split}
T'_{H,\kappa}\ : & \ (\zeta^{\Lambda}, \tilde\zeta_\Lambda, \sigma)\mapsto \big(\zeta^\Lambda + \eta^\Lambda,
 \tzeta_\Lambda+ \tleta_\Lambda , \\
& \
\sigma\mapsto \sigma + 2 \kappa
- \tleta_\Lambda \zeta^\Lambda+\eta^\Lambda \tzeta_\Lambda
+ 2\cR{H}\big)\, .
\end{split}
\ee
 It should be stressed that the extra shift
of $\cR{H} $ is crucial for the consistency of this action, since one finds that
$T'_{H_2,\kappa_2} \, T'_{H_1,\kappa_1} =
T'_{H_1+H_2,\kappa_3}$
where
\be
\kappa_3=
\kappa_1+\kappa_2+ \cR{H_1}+\cR{H_2}+\frac12  \langle H_1,H_2\rangle
-\cR{H_1+H_2}
\ee
is an integer, by virtue of \eqref{qrifprophol}. The extra shift in $\sigma$
was recently observed in the context of rigid CY compactifications
upon assuming invariance under a certain natural arithmetic group \cite{Bao:2010cc},
and, with hindsight, could also have been detected in a similar construction in the
non-rigid case \cite{Pioline:2009qt}.

At this point, we can now explain why the equality of the two quadratic
refinements $\sigma_{\text{D}}$ and $\sigma_{\text{NS5}}$ is desirable.
On the one hand, using \eqref{thschar}, one sees that a change $\Theta\mapsto \Theta'$
of the characteristics governing  fivebrane instanton
corrections leaves  \eqref{NS5coupling} invariant provided one  redefines the axions into
\be
\label{redef}
\hat C=C+\Theta-\Theta'\, ,
\qquad
\hat\sigma=\sigma+\langle \Theta-\Theta', C \rangle - \langle \Theta, \Theta' \rangle\, .
\ee
On the other hand, a change $\Theta_{\text{D}}\mapsto \Theta'_{\text{D}}$
of the characteristics $\Theta_{\text{D}}$ governing the D-instanton contributions
leaves \eqref{d2quali} invariant provided it is accompanied by a similar field
redefinition \eqref{redef} where $\Theta,\Theta'$ are replaced
by $\Theta_{\text{D}},\Theta'_{\text{D}}$. These two field redefinitions
are compatible as long as $\Theta-\Theta_{\text{D}}=\Theta'-\Theta'_{\text{D}}$.
Thus, if one wants to ensure that physics (in particular the moduli space $\cM$)
is independent  of the choice of quadratic refinement, as is known to be
the case in $\cN=2$ gauge theories \cite{Gaiotto:2008cd,Gaiotto:2010be}, one
must require that the difference $\Theta-\Theta_{\text{D}}$ is fixed. Since
the difference of two characteristics transforms (modulo integers)
like a symplectic vector under monodromies, and since the monodromy group
in general does not admit any invariant symplectic vector, the most natural
choice is to set $\Theta=\Theta_{\text{D}}$. This equality is
also generally expected from S-duality, which relates NS5- and D5-brane
instantons on the type IIB side.

\section{Topology of the NS-axion circle bundle $\cC$ over $\cM_c(\cX)$}

\noindent  So far we have established the fibration of the NS-axion circle bundle $\cC$ over the
torus of $C$-fields. The next question is to understand the fibration of this bundle over
the complex structure moduli space $\cM_c(\cX)$. This  is obviously tied
with the metric-dependent normalization factor $\cF$ in the fivebrane  partition function,
which is notoriously subtle \cite{Moore:2004jv}.
We shall limit ourselves to some preliminary comments in this direction.

 To this aim, let us return to the
perturbative metric \eqref{hypmetone}, and compute the curvature of the connection \eqref{Dsigone} on the circle bundle $\cC$. Taking into account that $\sigma$
has periodicity two, and the value \eqref{cchi} of the one-loop parameter $c$, we then find
\be
\label{c1C}
\de \left(\frac{D\sigma}{2}\right) = \omega_\cT + \frac{\chi(\cX)}{24}\, \omega_\sk\, ,
\ee
where $\omega_\cT$ is, as before, the \kahler form on $\cT$, and
$\omega_\sk=-\frac{1}{2\pi}\de\cA_K$ is the \kahler form on the complex structure moduli space
$\cM_c(\cX)$. The first term in \eqref{c1C} indeed confirms our identification of $\cC\vert_{\cT}$
with the fivebrane circle bundle $\cC_{\Theta}$.

The second term in \eqref{c1C} suggests that $e^{\I\pi\sigma}$ is a section
of $\cL^{\frac{\chi(\cX)}{24}}$, where $\cL$ is the Hodge line bundle over $\cM_c(\cX)$, i.e. the bundle
whose sections transform as $s\mapsto s\, e^f$ under rescalings $\Omega_{3,0}\mapsto e^f
\Omega_{3,0}$ of the holomorphic 3-form on $\cX$. This however causes trouble, since
$\chi(\cX)$ is rarely a multiple of 24, and $\cL$ does not admit any natural 24:th root.
In particular, $f_M$ in Eq. \eqref{sigmon} is only defined modulo $2\pi\I$, which implies
that the shift of $\sigma$ under monodromies is ambiguous modulo $\chi(\cX)/12$.  Since
fivebrane charge quantization requires that $\sigma$ is periodic modulo 2,
there must be an additional constant shift of $\sigma$ in Eq. \eqref{sigmon}
 to resolve this  ambiguity.
A similar problem afflicts the topological string amplitude of the B-model on $\cX$,
which is claimed to transform as a section of $\cL^{\frac{\chi(\cX)}{24}-1}$
under monodromies \cite{Bershadsky:1994cx}. A proper understanding
of the topological nature of either the axion $\sigma$, the topological string amplitude
or the fivebrane partition function
presumably involves determinant line bundles, along the lines of
\cite{Belov:2006jd,Belov:2006xj}, and should allow to compute  the variation of the NS-axion
under monodromies.
We hope to address this issue in future work.

\section*{Acknowledgments}

\noindent  We are grateful to A. Kleinschmidt, J. Manschot, R. Minasian, S. Monnier,
B. Nilsson, C. Petersson,
Y. Soibelman, S. Vandoren and especially to G. Moore
for valuable discussions and correspondence. S.A. is grateful to Perimeter Institute for Theoretical Physics
for the kind hospitality and financial support during the course of this work. D.P. is grateful
to LPTHE Jussieu for hospitality where part of this work was carried out.

\end{document}